\documentclass[12pt]{iopart}
\usepackage{times}
\usepackage{graphicx}
\usepackage{amssymb}
\def\be{\begin{eqnarray}}
\def\ee{\end{eqnarray}}
\def\ba{\begin{array}}
\def\ea{\end{array}}

\begin{document}

\title[Thomas-Fermi-Poisson theory of screening in strong magnetic
fields]{Nonlinear Thomas-Fermi-Poisson theory of screening for a Hall bar 
under strong magnetic fields}

\author{A. Siddiki and Rolf R. Gerhardts}
\address{Max-Planck Institut f\"ur Festk\"orperforschung,
Heisenbergstrasse 1, D-70569 Stuttgart, Germany}%

\begin{abstract}
Low-temperature screening properties of the inhomogeneous two-dimensional
electron gas in a Hall bar subjected to a strong
perpendicular magnetic field are explored using
a self-consistent  approach. An external oscillating
modulation potential with an amplitude of the order of the cyclotron energy
is added to the electron-confining background potential, and the resulting
change of the self-consistent potential is investigated as function of
modulation strength, magnetic field, 
and temperature. The consequences of Landau-level pinning and the interplay of
compressible and incompressible regions for the resulting strongly non-linear
screening phenomena are explained.
\end{abstract}
% **** Leave the next line commented out! ****

% \maketitle

\section{Introduction}
As a consequence of the highly degenerate, Landau-quantised energy levels, 
a two-dimensional electron gas (2DEG) in a strong perpendicular magnetic 
field
has unusual low-temperature screening properties
\cite{Wulf88:4218,Efros88:1019}. In an inhomogeneous 2DEG with long-range
density fluctuations, pinning of Landau levels (LLs) leads
to quasi metallic regions with high density of states (DOS) at the Fermi
energy and therefore nearly perfect screening ability. These so called
``compressible'' regions coexist with quasi-insulating
``incompressible'' regions separating compressible regions with different 
LLs at the Fermi energy. In the incompressible regions the Fermi energy falls
into the gap between two adjacent LLs and, in thermodynamic equilibrium at low
temperatures, no redistribution of electrons is possible which could
contribute to screening. Also, in the incompressible
regions the electron density $n_{el}({\bf r})$ is constant, corresponding to
total filling of an integer number of LLs, while in the compressible regions
$n_{el}({\bf r})$ adjusts itself so that the self-consistent electrostatic 
potential energy $V({\bf r})$ of an electron differs from the Fermi energy,
more precisely from the constant electrochemical potential $\mu^{\star}$ by a
Landau energy $\hbar \omega_c 
(n+1/2)$, where $\omega_c=eB/m$ is the cyclotron frequency in the magnetic
field $B$. As a consequence, $V({\bf r})$ differs between different
compressible regions by integer multiples of  $\hbar \omega_c$. 
Landau level pinning and the interplay of
compressible and incompressible regions will strongly affect the screening
properties of the 2DEG.

In a bounded 2D geometry with translation invariance in one direction, the
compressible and incompressible regions degenerate to strips parallel to the
boundary. For half-space and Hall-bar geometries models with planar charge
distributions have been proposed
that allow closed solutions of Poisson's equation, i.e., the
calculation of the potential for given electron density,  and
estimates of position and widths of the incompressible strips have been given
 \cite{Chklovskii92:4026,Chklovskii93:12605}. By adding  the
non-linear Thomas-Fermi approximation for the calculation of the electron
density from the potential,
 that work was  extended to a self-consistent approach
which allows to calculate both electron density and electrostatic potential for
arbitrary temperature \cite{Lier94:7757,Oh97:13519}. This approach shows that
the existence and the width of 
incompressible strips depends sensitively on temperature. It also permits 
to calculate their position and width for given background charges without
additional assumptions. We adopt this approach to investigate systematically
the screening of a harmonic external potential in such a confined 2DEG.

%%%%%%%%%%%%%%%%%%%%%%%%%%%%%%%%%%%%%%%%%%%%%%%%%%%%%%%%%%%%%%%%%%%%%%%%
\section{The Thomas-Fermi-Poisson (TFP) approach}
%%%%%%%%%%%%%%%%%%%%%%%%%%%%%%%%%%%%%
In order to employ the approach presented in Ref.~\cite{Oh97:13519}, we assume
that our  2DEG lies in the $z=0$ plane, laterally
confined by in-plane gates located at $x<-d$ and $x>d$, which are kept at
voltages $V_{L}$ and $V_{R}$, respectively. In this work we assume a symmetric
depletion on both sides with the depleted strips having a width of 
$|d-b|$, and we will consider only symmetric potentials $V(-x)=V(x)$, notably
with $V(\pm d)=V_L=V_R=0$,  which
leave the electron density profile symmetric, $n_{el}(-x)=n_{el}(x)$.
Then, solution of Poisson's equation yields for the electrostatic potential
energy of an electron
\be  
\hspace*{-2cm}
V(x)=- \frac{2e}{\bar{\kappa}} \int_{-d}^{d}\!dt K(x,t)\,\rho(t), 
\qquad K(x,t)= \ln
\bigg|\frac{\sqrt{(d^{2}\!-\!x^{2})(d^{2}\!-\!t^{2})}+d^{2}\!-\!tx}{(x-t)d}
\bigg|,
\label{intpoten}
\ee
where $\rho(x)$ is the surface charge density in the strip $|x|<d$ and
$\bar{\kappa}$ is an average dielectric constant
\cite{Oh97:13519}. If we assume a 
homogeneous surface density $n_0$ of positive background charges,
$\rho(x)=e[n_0-n_{el}(x)]$, we obtain 
\be V(x)=V_{bg}(x)+V_H(x), \quad \mbox{where} \quad
V_{bg}(x) =-E_{0}\sqrt{1-(x/d)^{2}} \label{potsum}
\ee
 with $ E_{0}=2\pi e^{2}n_{0}d/\bar{\kappa}$
is the confining potential due to the background and
$V_H(x)$ the Hartree contribution due to the electron density, given by
Eq.~(\ref{intpoten}) with $\rho(x)$ replaced by $-e n_{el}(x)$.
Assuming that  $V(x)$ varies on a characteristic length much larger than
typical quantum lengths, notably the magnetic length
$l_{m}=\sqrt{(\hbar/eB)}$, one can calculate $n_{el}(x)$ for given $V(x)$ in
the Thomas-Fermi  approximation \cite{Oh97:13519}
\be \hspace*{-1cm}
 n_{el}(x)=\int dED(E)f\big( [E+V(x)-\mu^{\star}]/k_{B}T \big), \quad \mbox{with}
\quad f(\epsilon)=[1+e^{\epsilon}]^{-1}
\label{tomf}\ee
 the Fermi function, $\mu^{\star}$ the 
electrochemical potential and $T$ the temperature.
Here $D(E)$ is the density of states of the 2DEG,   
which we take to be the bare spin-degenerate Landau DOS,
\be D(E)=(\pi l_{m}^{2})^{-1}
\sum_{n=0}^{ \infty} {\delta\big(E-\hbar\omega_{c}(n+1/2)\big)}
\label{dos}.\ee
This completes the TFP scheme for the self-consistent
calculation of $n_{el}(x)$ and $V(x)$ for given temperature and magnetic field
and homogeneous background charge. To study screening effects, we add to
$V_{bg}(x)$ in Eq.~(\ref{potsum}) the harmonic external potential
\be V_{m}(x)=V_{0} \cos(k_{\lambda}x), \quad \mbox{with} \quad
 k_{\lambda}=(\lambda+1/2) \pi/d, \label{modpot}
\ee
where $\lambda$ is an integer (to preserve the boundary conditions),
calculate the resulting self-consistent potential and electron density as
 function of the  modulation strength $V_0$, and compare these results with
 those for $V_0=0$.  [In principle $ V_{m}(x)$ can be generated by a
 redistribution of the background charge density.]

For technical reasons we always start the iterative solution of the
self-consistent TFP scheme with a calculation for vanishing temperature and
magnetic field and for homogeneous background, i.e., we take
$ n_{el}(x)=D_{0}[E_{F}- V(x)]\theta \big(E_{F}- V(x)\big) $ with $D_{0}=m/
\pi \hbar^{2}$ as 2D DOS and define the symmetric density profile by the
requirement $V(b)=V(-b)=E_F=\mu^{\star}(T=0)$.  Then Eq.~(\ref{intpoten}) reduces to a
linear  integral equation,
\be V(x)=-  
E_{0}\sqrt{1-\left(\frac{x}{d}\right)^{2}}+ \frac{1}{\pi
  a_{0}}\int_{-b}^{b}\! dt \big[{E}_{F} - V(t)\big]\,K(x,t) \label{sikim} \ee
with  $a_{0}=\bar{\kappa}/2 \pi e^{2}D_{0}$ the bare 
screening length. [Due to a misprint, in the corresponding Eq.~(15) of
Ref.~\cite{Oh97:13519} the factor $\pi$ in the denominator is missing.]
The choice of $b$ defines  the
average density, which is then kept fixed for the calculations at 
finite
temperature, magnetic field, and modulation.

%%%%%%%%%%%%%%%%%%%%%%%%%%%%%%%%%%%%%%%%%%%%%%%%%%%%%%%%%%%%%%%%%%%%%%
\section{Results and discussion}
%%%%%%%%%%%%%%%%%%%%%%%%%
%
In the numerical calculations we keep some parameters fixed. So we divide the
interval $-d<x<d$ into $500$ subintervals and calculate $V(x)$ and
$n_{el}(x)$ 
on the corresponding equidistant mesh points $x=x_n$. The depletion length is
defined by taking $b/d=0.9$ and the bar width by $\pi a_0/d=0.01$. For GaAs
values ($\bar{\kappa}=12.4$) $a_0\approx 5\,$nm, this means $2d\approx
3\,\mu$m, which is small compared to typical Hall bars used in
experiments. Calculation for smaller $a_0/d$, i.e., larger $d$ requires 
more subintervals, and thus larger storage and computation
time, to achieve the same accuracy near the incompressible strips, but yields
no qualitative changes. Therefore, we fix the mentioned parameter values and
obtain from the solution of the system of linear equations defined by 
Eq.~(\ref{sikim}) an average electron density that yields the average LL
filling factor $\bar{\nu}=2$ for a cyclotron energy $\hbar\omega_c 
=\Omega_2\equiv 0.2311 \times 10^{-2} E_0$. Starting from this solution we
find the solution of the non-linear TFP scheme at finite $T$ and $B$
implementing the Newton-Raphson method \cite{Lier94:7757,Oh97:13519}.  
For the bare modulation potential (\ref{modpot}) we take $\lambda=2$, so that
$V_m(x)$ has a maximum at $x=0$ and two minima at $x=\pm 0.4 d$, i.e. well 
inside
the Hall bar, whereas near the maxima at $x=\pm 0.8 d$ the confinement
potential increases strongly and the electron density starts to decrease,
so that the screening properties are obscured by boundary effects.

\begin{figure} $ \left. \right. $ \includegraphics{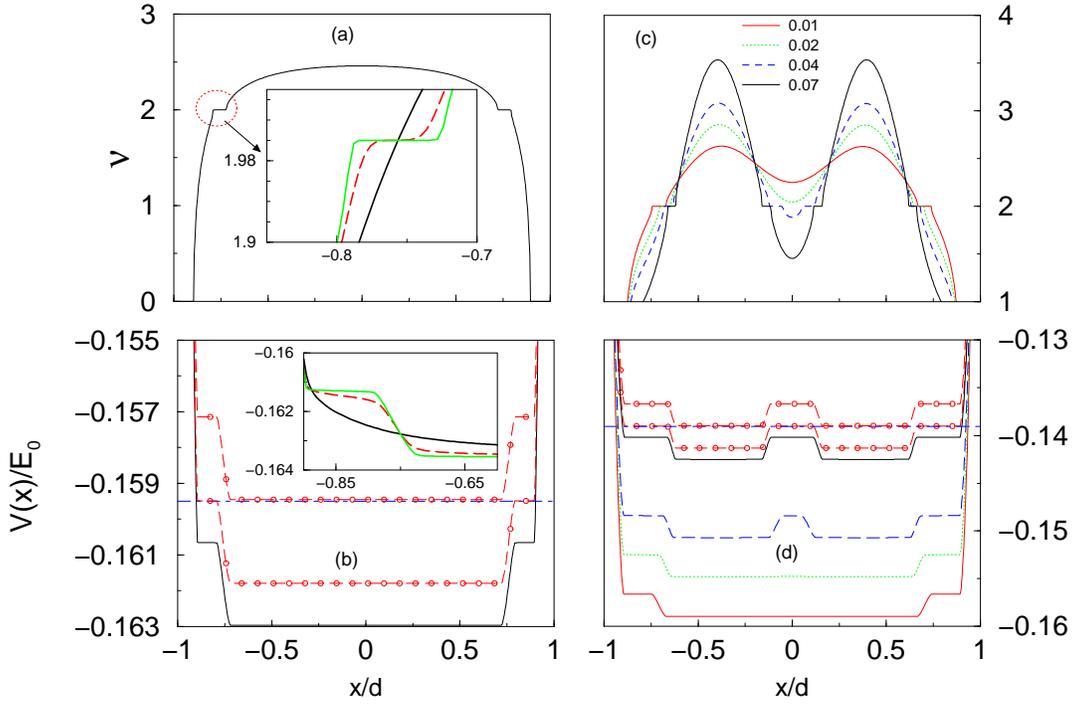} \vspace{8.7cm}
\caption{Electron densities [(a),(c)] and electrostatic
potentials $V(x)$ [(b),(d)] without modulation [(a),(b)] and with 
external
modulation potential $V_{m}(x)=V_{0}\cos(qx)$, $q=2.5\pi/d$  
[(c),(d)], where $V_{0}/E_{0}$ values are given by the legend. 
Insets in (a) and (c) demonstrate the temperature dependence for
$kT/E_{0}=0.1\times10^{-3}$ (thick solid line), $0.2\times10^{-4}$ (dashed
line), and $0.1\times10^{-4}$ (thin solid line).}
\label{fig1}
\end{figure}

Figure \ref{fig1} shows the potentials and electron densities of the Hall 
bar calculated for several modulation amplitudes at strong magnetic fields 
and low temperatures. On the left panel we consider the case $V_{0}=0$ and 
show the 
temperature dependence in the inset, while the right panel shows the 
evolution of the potential and electron density with increasing modulation 
amplitude. All the electron densities are 
expressed in terms of local filling given by $\nu(x)=2\pi 
n_{el}(x)l^{2}_{m}$ and potentials are normalised by $E_{0}$. 

 Figure \ref{fig1}a presents the 
electron density, with two incompressible strips located symmetrically near
$x/d\,$=$\, \pm0.75$ and the surrounding compressible regions. With
decreasing temperature the incompressible strips are  stronger pronounced (see
inset Fig.~\ref{fig1}a) and  so are the related potential steps (inset
Fig.~\ref{fig1}b). As mentioned in the introduction, in compressible regions
 one of the LLs is pinned at $\mu^{\star}$, leading to
 nearly  perfect screening and constant potential, while in an incompressible
 region $\mu^{\star}$ falls into a gap between the LLs making redistribution
 of  electrons, and hence screening, energetically impossible and  $n_{el}(x)$
constant. 
We demonstrate these features in Fig.~\ref{fig1}b, where we exhibit
$\mu^{\star}$ (dash-dotted line), the first two LLs (dashed lines with opaque
circles),  and the total potential (solid line).

 In the right panels we show  
electron density and potential profiles for various modulation amplitudes
$V_0$ at a low temperature and a high magnetic field, at which the average
filling factor in the centre region is around 2.5, i.e., the  $n=1$ LL is
pinned to $\mu^{\star}$. With increasing $V_0$ the
external potential (energy) is raised in the centre and lowered near
$x/d\,$=$\, 
\pm0.4$, which leads to a lowering of $n_{el}(x)$ near $x=0$ and an increase
near $x/d\,$=$\,\pm0.4$. This redistribution of charges screens the potential
at the low temperature of  Fig.~\ref{fig1}c,d so effectively, that the resulting
weak modulation of the self-consistent potential for $V_0/E_0=0.01$ and 0.02
is not seen on the scale of  Fig.~\ref{fig1}d. At larger modulation, 
however, the filling
factor $\nu(0)$ in the centre falls below 2, which means that near $x=0$ 
the
self-consistent potential must rise so much that the lowest LL $n=0$ reaches
$\mu^{\star}$ (as is demonstrated for  $V_0/E_0=0.07$ in Fig.~\ref{fig1}d). 
Therefore, on both sides of the centre region incompressible strips must
develop across which the self-consistent potential changes by
$\hbar\omega_c$. 

We see from these results that the screening is highly non-linear, since 
the
change of the self-consistent potential produced by the harmonic modulation
(\ref{modpot}) is step-like rather than cosine-like. Moreover, for weak
modulation $V_0/E_0 \lesssim 0.02$
the resulting variation $\Delta V=V(0)-V(0.4d)$ of the self-consistent
potential is so small that it is not detectable on the scale of
Fig.~\ref{fig1}d, whereas $\Delta V$ for  $V_0/E_0 \gtrsim 0.04$ assumes the
constant value $\hbar\omega_c$. Apparently this stepwise increase of $\Delta V$
with $V_0$ is a consequence of the pinning of LLs to the electrochemical
potential (locally perfect screening), which is stronger pronounced at 
lower temperatures.
We now consider this effect in more detail.

\begin{figure} $ \left. \right. $ \includegraphics{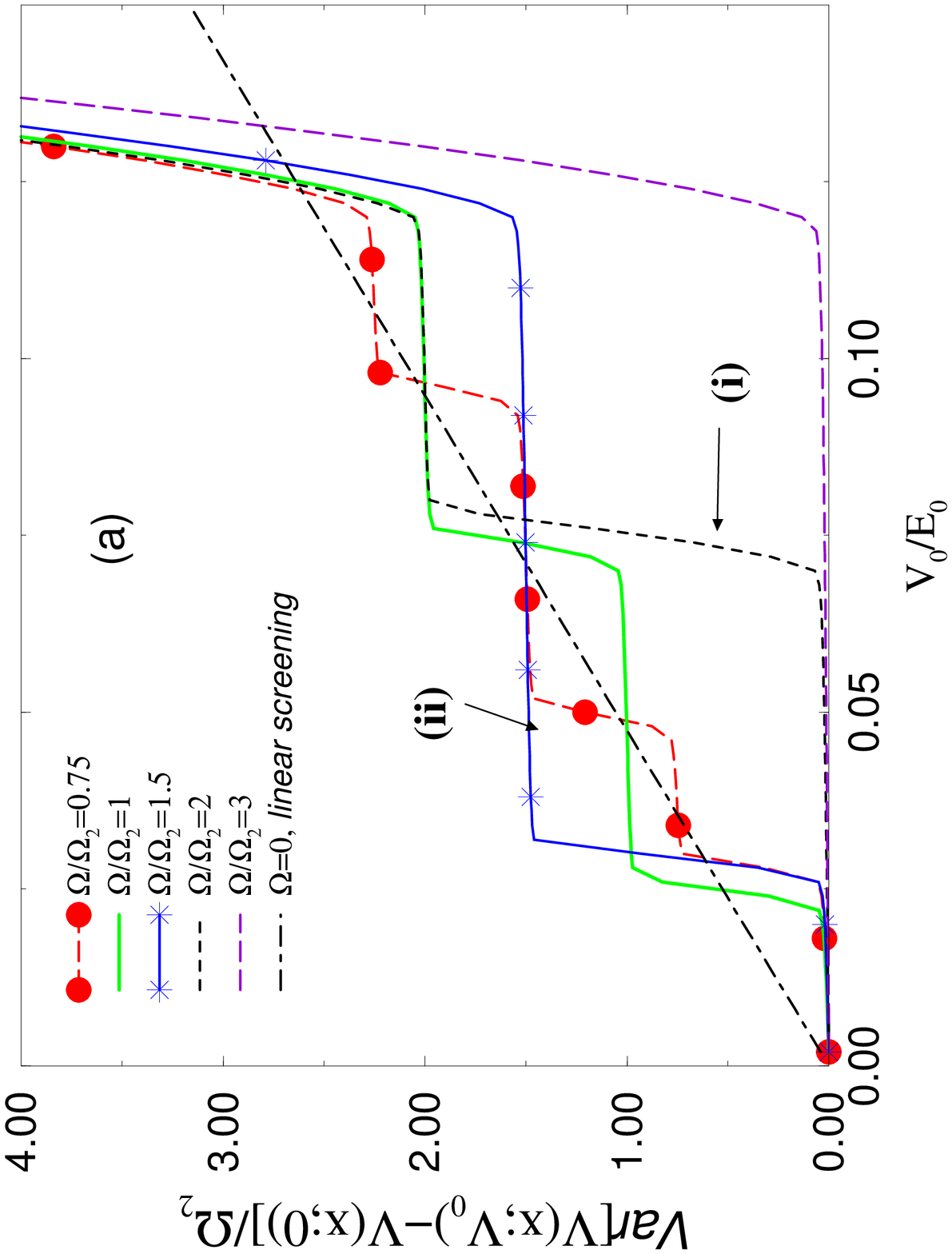}
$ \left. \right. $ \includegraphics{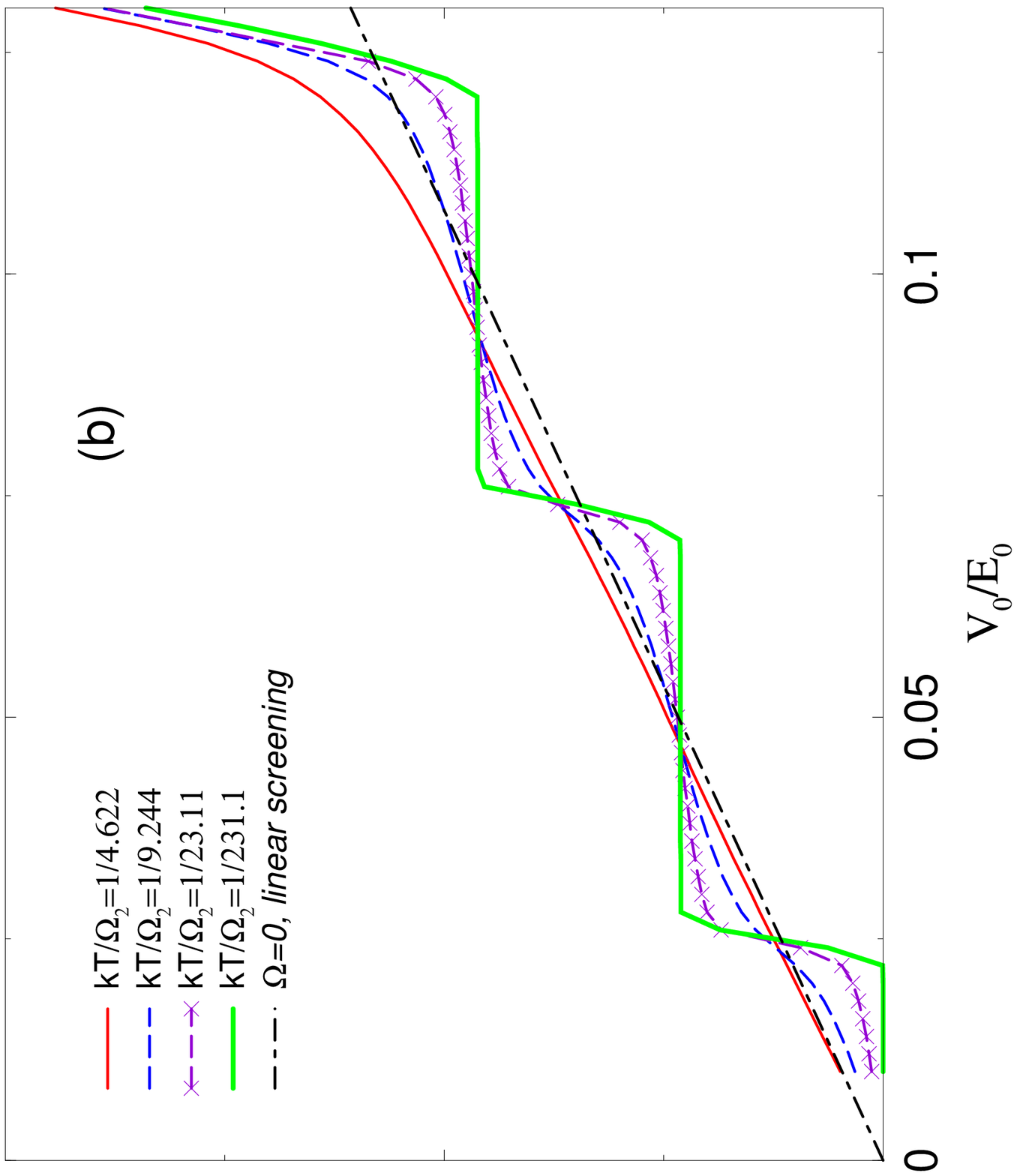} \vspace{5.5cm}
\caption{(a) Variance of screened potential versus amplitude $V_{0}$
of the external potential for various magnetic field values
($\pi a_0/d=0.01$, $k_BT/E_0=2\cdot 10^{-5}$). The
dash-dotted line indicates linear screening for zero $B$ and $T$, see text.
 (b) Temperature dependence of the variance for $\Omega/\Omega_{2}=1$}.
\label{fig2}
\end{figure}
%
%%%%%%%%%%%%%%%%%%%%%%%%%%%%%%%
%%%%%%%%%%%%%%%%%%%%%%%%%%%%%%%
\subsection{Tuning the modulation strength at fixed magnetic fields }
%%%%%%%%%%%%%%%%%%%%%%%%%%%%%%%
%
In Fig.~\ref{fig2}a we show the ``variance'' of the screened potential as
function of the amplitude $V_0$ of the applied modulation for several values
of the magnetic field and a fixed low temperature, and in Fig.~\ref{fig2}b the
same dependence for a fixed value of the cyclotron energy and several
temperatures.  We define this variance as the difference between the maximum
(at $x=0$) and the minimum (near $x=0.4d$)  of the difference 
$V(x;V_0)-V(x;0)$ of the self-consistent potentials calculated with and
without modulation, respectively.

For a fixed value of the magnetic field (i.e., of $\Omega\equiv\hbar
\omega_c$), we obtain a stepwise increasing curve, as we expect from the
discussion of   Fig.~\ref{fig1}c,d, which corresponds to 
$\Omega=\Omega_2$. The
corresponding (thick solid) curve in Fig.~\ref{fig2}a shows two broadened steps
around $V_0/E_0=0.025$ and 0.075. We have already discussed the reason for the
step at the lower $V_0$ value and for the plateaus below and above this
step. With further increasing $V_0$ the filling factor near the density maxima
becomes larger than 4 (see Fig.~\ref{fig1}c) while the density in the centre is
still finite. Then new incompressible strips with local filling $\nu(x)=4$
develop on both sides of the density maxima, which corresponds to a 
decrease of
the self-consistent potential by the amount $\Omega$ to  new local minima
near  $x=\pm0.4d$. In Fig.~\ref{fig2}a this leads to the plateau of height
$2\Omega$. As $V_0$ increases further, the electron density $n_{el}(0)$ in the
centre vanishes and no further redistribution of electrons from the centre
region to the region of maximum density is possible. Then the high screening
ability of the electron system breaks down and 
the variance increases with further increasing  $V_0$ much more rapidly (see
behaviour at $V_0/E_0>0.125$). This breakdown happens for $\Omega=\Omega_2$
when the filling factor at the density maxima is somewhat larger than 4
($\sim5$).  

We can now understand all the curves in Fig.~\ref{fig2}a. Since, apart 
from the
fine structure near incompressible strips, the global density profile depends
only weakly on the magnetic field, for the lowest curve
(long-dashed, $\Omega=3\Omega_2$) typical filling factors $\nu(x)$ are by a
factor 3 smaller than for the case $\Omega=\Omega_2$ we have just
discussed. Therefore the breakdown situation is 
reached when the filling factor near the density maxima is of the order
$5/3<2$. Thus the breakdown sets in before an incompressible strip and thus a
plateau of height $3\Omega_2$ of the variance can develop.
For $\Omega=2\Omega_2$ [short-dashed line, label (i)] the local filling
factors $\nu(x)$ are typically about half as large as in the case
$\Omega=\Omega_2$. Thus, $\nu(x)<2$  for weak modulation, and incompressible
strips with  $\nu(x)=2$ can occur only near the density maxima for
sufficiently strong modulation, and a single plateau of height $2\Omega_2$
occurs below the breakdown regime. A similar situation is met for
$\Omega=1.5\Omega_2$ [label (ii)], but since now  $\nu(x)$ is about $2/3$
times that for $\Omega=\Omega_2$, the threshold of the plateau is reached at a
considerably weaker modulation. Finally, for $\Omega=0.75\Omega_2$ we obtain
three plateaus in the pre-breakdown regime, corresponding to the successive
development of incompressible strips with local filling factor 4 (maxima), 2
(minimum), and 6 (maxima).

We should mention that the behaviour of the variance curves is not always that
regular. For special values of $\Omega$ it may happen that incompressible
strips near the centre and near the density maxima occur at the same
modulation strength. Then steps of height $2\Omega$ occur. We met such a
situation for $\Omega=0.5\Omega_2$ (not shown in Fig.~\ref{fig2}a), which
yields nearly the same trace as $\Omega=\Omega_2$.

For the smaller $\Omega$ values, the variance curves oscillate around the
dash-dotted line in   Fig.~\ref{fig2}a, which represents the corresponding
result  $2V_0/[E_0 \varepsilon(q)]$ of linear Thomas-Fermi screening
\cite{Wulf88:4218} for zero $T$ and $B$, with
$\varepsilon(q)=1+1/(a_0q)=41$ for $q=2.5 \pi/d$.
This linear screening approximation breaks down when the amplitude of the
screened potential (i.e. half the variance) becomes equal to the Fermi energy
of the unmodulated system \cite{Wulf88:162}, which can be estimated from 
 Fig.~\ref{fig1} as $\sim 3\times 10^{-3}E_0$, corresponding to $V_0 \sim 0.12
 E_0$. Finally we should mention that the small-variance regime in
 Fig.~\ref{fig2}a also corresponds to linear screening, but now with
 $\varepsilon(q)=1+(D_T/D_0)/(a_0q)$, where $D_T$ is the thermodynamic DOS
 in a LL, which can be estimated by $D_T/D_0 \sim \hbar\omega_c/(4k_BT)$
 \cite{Wulf88:4218}. Thus, the slope of the variance curves is by a factor
 $\sim 4k_BT/\Omega \ll 1$ smaller than that of the dash-dotted line.

As shown in Fig.~\ref{fig2}b, with increasing temperature the plateaus of the
variance-versus-$V_0$ curves get a finite slope and the steps are smeared
out. This is an immediate consequence of the temperature dependence of the
incompressible strips (see Fig.~\ref{fig1}a). At the highest shown temperature
screening becomes again linear and independent of $B$, but now the temperature
is so high that the thermodynamic DOS is noticeably smaller than $D_0$. A
clear indication of steps is seen only for $k_BT/\hbar \omega_c\lesssim 1/25$
\cite{Lier94:7757}. 

%%%%%%%%%%%%%%%%%%%%%%%%%%%%%%%%%%%%%%%%%%%%%%%%%%%%%%%%%%%%
%%%%%%%%%%%%%%%%%%%%%%%%%%%%%%
\subsection{Sweeping the magnetic field at fixed modulation strength}
%%%%%%%%%%%%%%%%%%%%%%%%%%%%%%%%%
%
From the above considerations we expect that, for fixed modulation strength
$V_0$, the variance as a function of the cyclotron energy $\Omega$ will mainly
follow integer multiples of $\Omega$ with abrupt transitions between
neighbouring integers corresponding to the steps in  Fig.~\ref{fig2}a. These
abrupt transitions will be similar to the  familiar magnetic-field-dependent
jumps of the chemical potential of a homogeneous 2DEG at fixed density, which
have nothing to do with the periodic modulation and should be separated from
the modulation-induced structures. To this end, we first discuss the 
chemical
potential. 

\begin{figure} $ \left. \right. $ \includegraphics{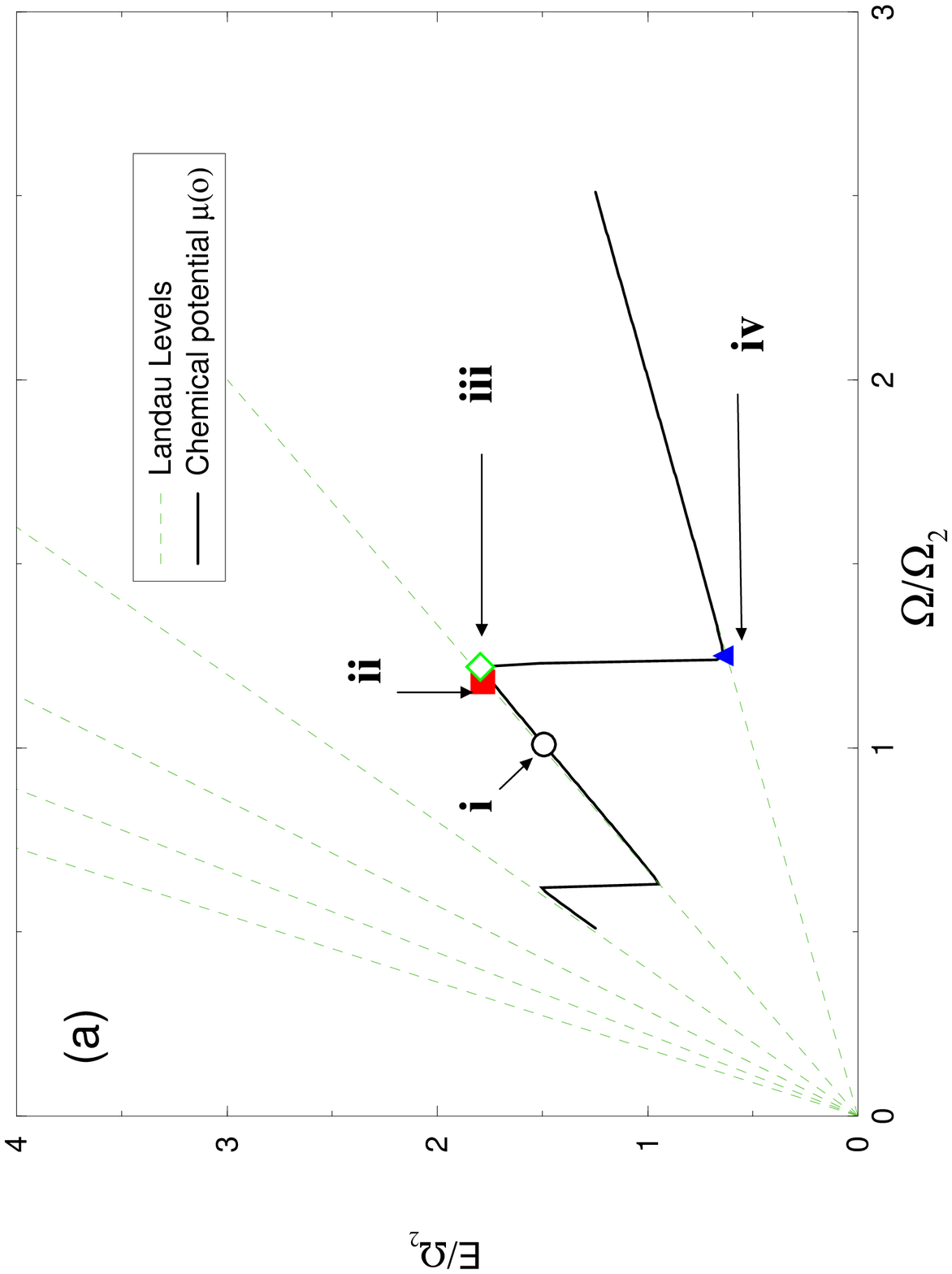} \vspace{10.0cm}
$ \left. \right. $ \includegraphics{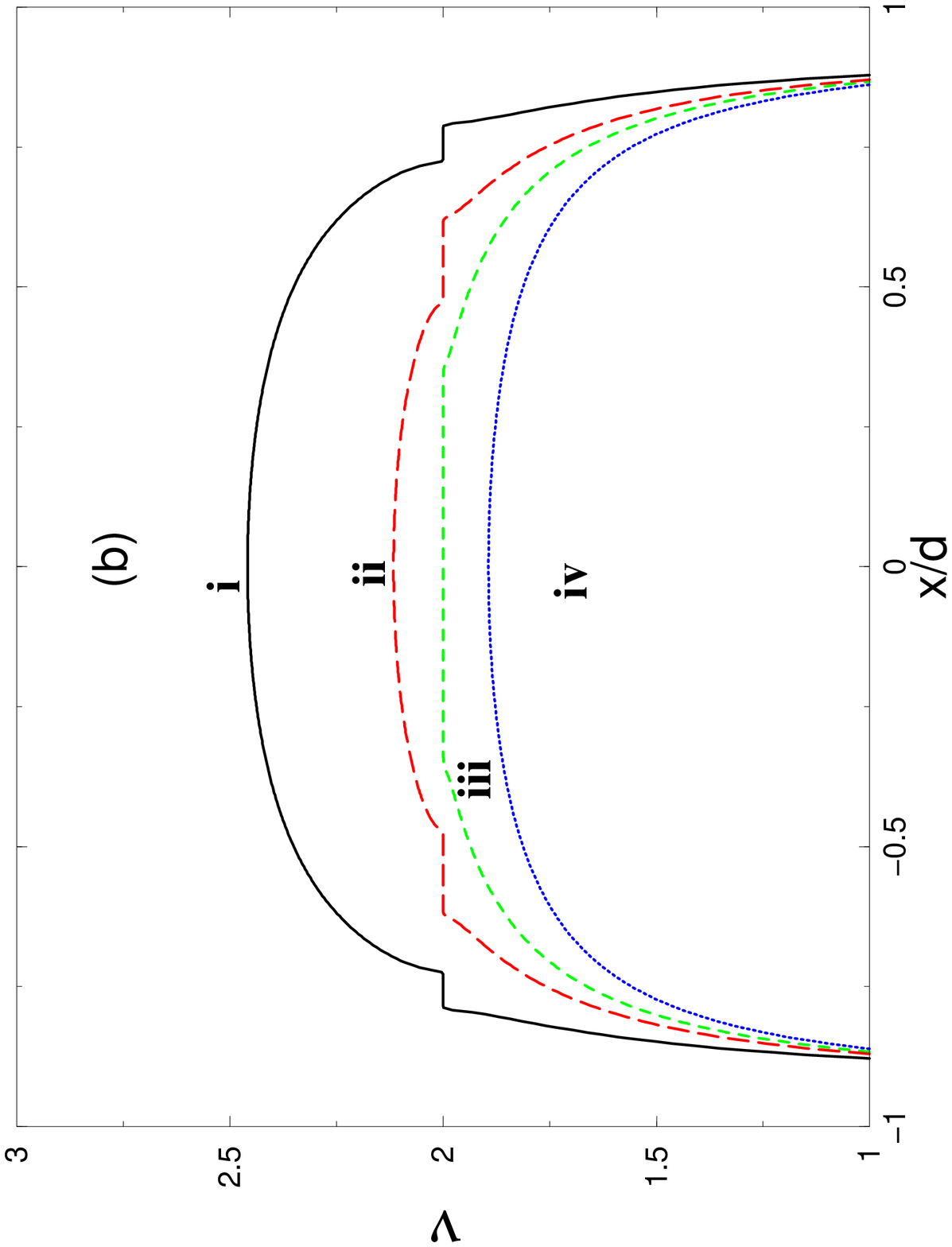}
$ \left. \right. $ \includegraphics{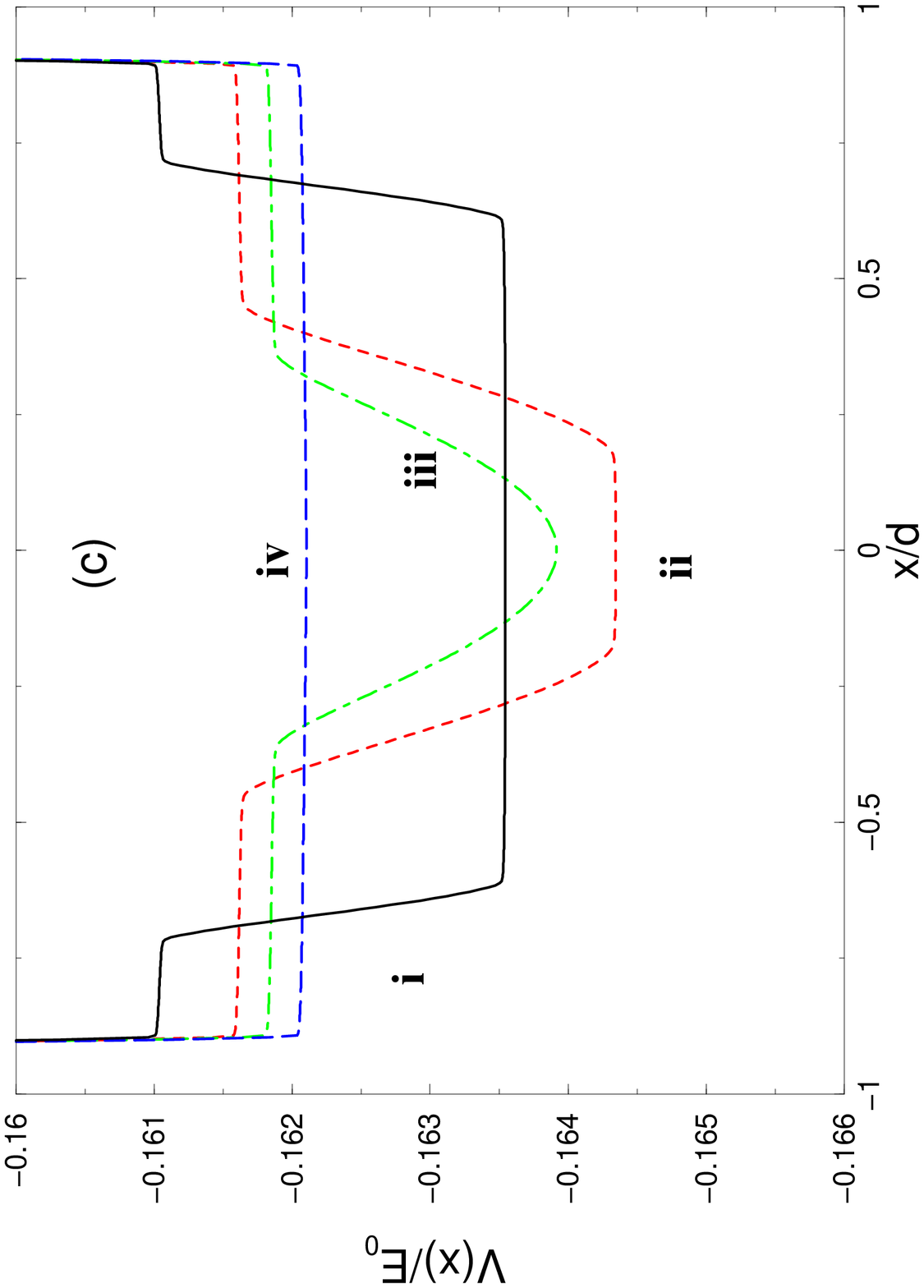}
\caption{Chemical potential and LLs (a); selected density (b)
  and potential (c) profiles.}
\label{fig4}
\end{figure}

In a Hall bar with spatially varying electron density and electrostatic fields
 the electrochemical potential $\mu^{\star}$ is a thermodynamic quantity
 being constant in equilibrium. As an analogue of the chemical potential
 of a homogeneous 2DEG we define $\mu(x)=\mu^{\star}-V(x)$ and denote in the
 following $\mu(0)$ as the chemical potential. In Fig.~\ref{fig4}a we show,
 together with the LLs $\Omega(n+1/2)$, a
 plot of $\mu(0)$ versus $\Omega$, which exhibits a saw-tooth shape as
 known from the unbounded 2DEG. To understand this, we plot in
 Fig.~\ref{fig4}b and \ref{fig4}c filling factor and self-consistent potential
 $V(x)$ for four magnetic field values marked in Fig.~\ref{fig4}a.
For the lowest selected $B$ value [opaque circle in Fig.~\ref{fig4}a, label
(i)], in the centre region the LL with $n=1$ is pinned to $\mu^{\star}$, the
filling in the centre is $\nu(0)>2$, and the incompressible strips with
$\nu(x)=2$ are close to the edges. As $\Omega$ increases these strips move
towards the centre, and the compressible centre region, with  $\nu(x)>2$ and a
flat potential minimum, shrinks [case (ii), filled square]. In the transition
region at still larger $\Omega$  [(iii), opaque diamond] the strips merge to
an incompressible centre region, where the potential minimum is no longer
flat, and its depth (measured from the adjacent compressible strips) is 
smaller
that $\Omega$. As  $\Omega$ sweeps through the transition region, the width of
the incompressible  centre region and the depth of the potential minimum shrink
 to zero.  When the transition is completed [(iv), filled triangle], we have
 again a broad compressible centre region in which the next lower LL (here 
$n=0$)
 is pinned to $\mu^{\star}$. Thus, in the confined 2DEG in 
 a Hall bar, the jump of the chemical potential from a LL to the next lower
 one  is realised by a drastic change of the position dependence of the
 self-consistent potential, which in our idealised model happens in the
 centre.  
\begin{figure} $ \left. \right. $ \includegraphics{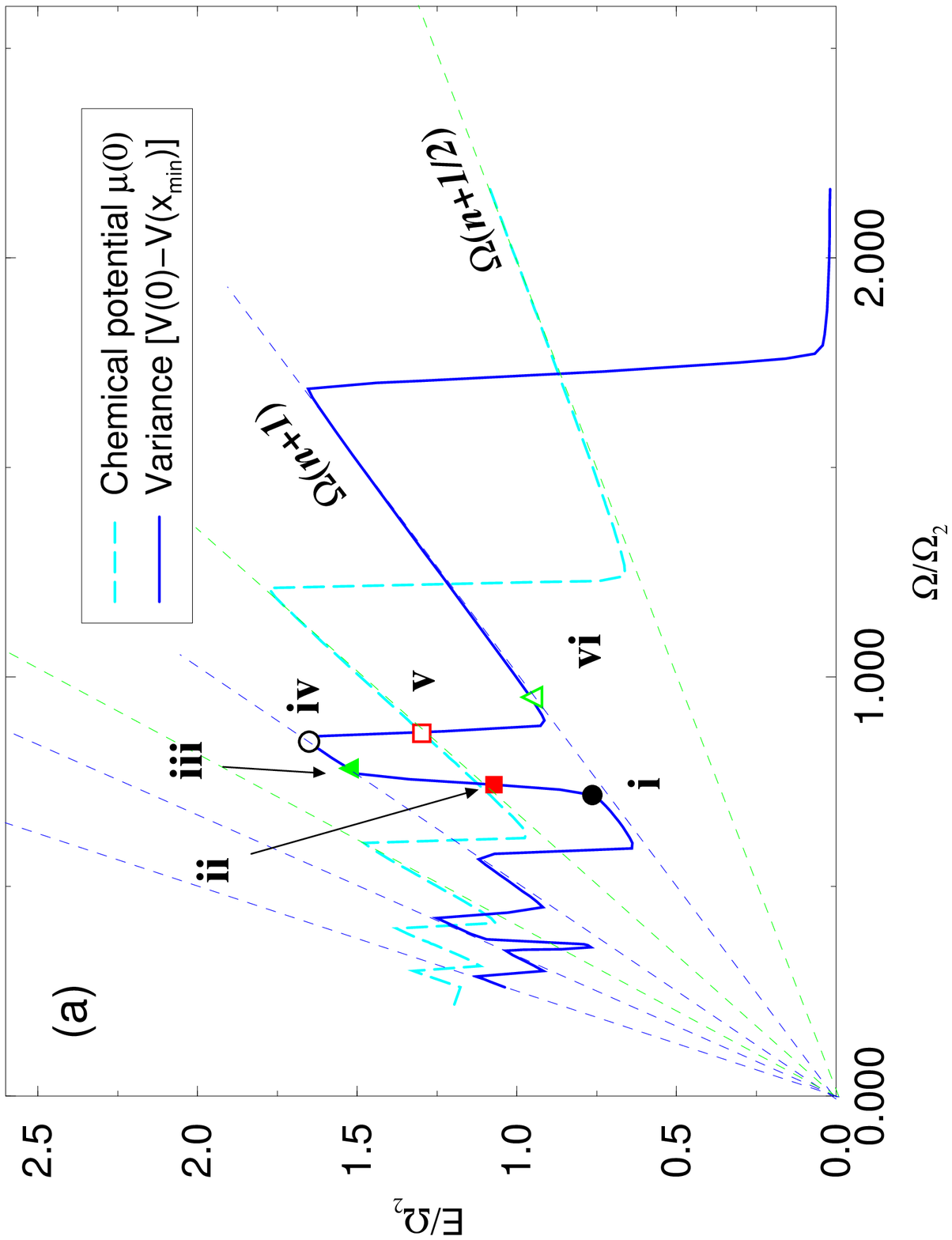} \vspace{11.5cm}
$ \left. \right. $ \includegraphics{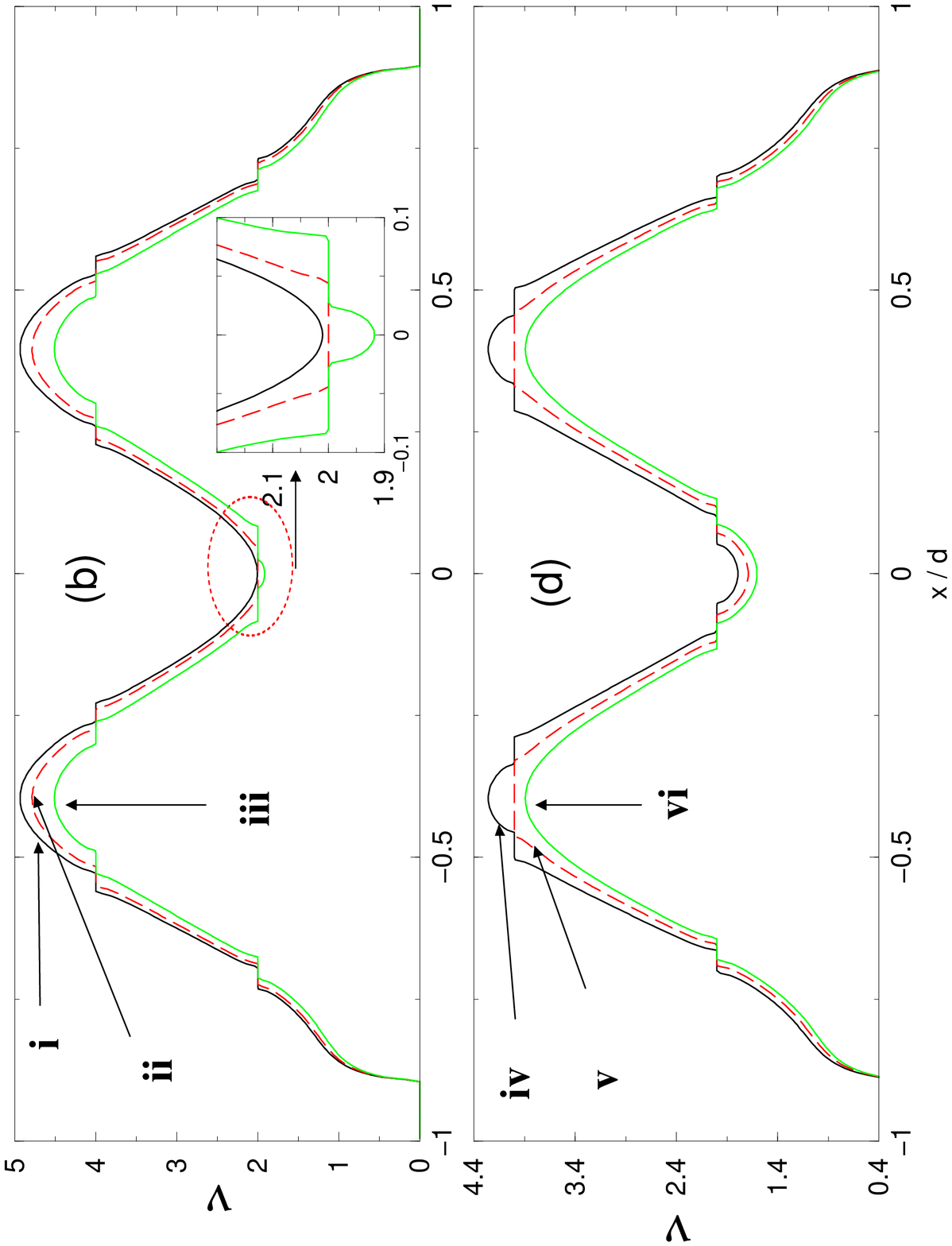}
$ \left. \right. $ \includegraphics{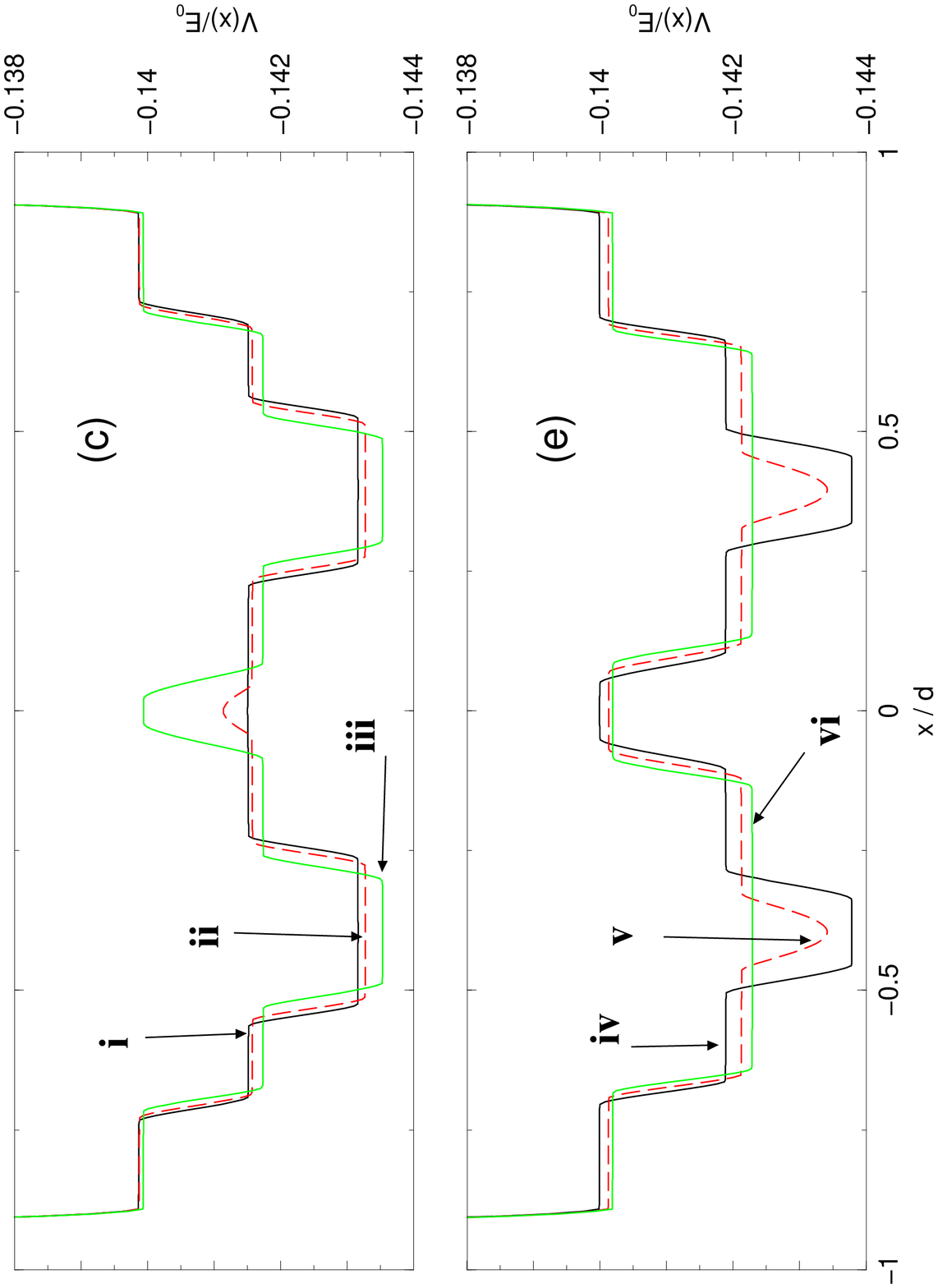}
\caption{(a) Variance of self-consistent potential for modulation amplitude
  $V_0=0.05E_0$ versus $\Omega$; (b),(d) density and (c),(e) potential
  profiles for selected  $\Omega$ values marked in (a). $k_BT/E_0=2\cdot
  10^{-5}$, $\pi a_0/d=0.01$.}
\label{fig5}
\end{figure}

To separate these effects from the modulation-induced screening effects, we
define now the variance of the self-consistent potential as
$var=V(0)-V(x_{min})$, where $\pm x_{min}\sim \pm 0.4d$ are the positions of
the minima of 
$V(x)$ in the presence of modulation, and $V(0)$ is the local maximum at the
centre. In Fig.~\ref{fig5}a we plot $var$ versus $\Omega$, together with
$\mu(0)$. As expected, $var$ follows widely integer multiples of $\Omega$,
whereas $\mu(0)$ follows half-integer ones. But, whereas $\mu(0)$ jumps with
increasing $\Omega$ always to the next lower LL, $var$ can also jump to a
higher multiple of $\Omega$. 
To understand this, we show in Fig.~\ref{fig5}b-\ref{fig5}e density and
potential profiles for the six $\Omega$ values marked in  Fig.~\ref{fig5}a.

In case (i) (solid lines in Fig.~\ref{fig5}b and \ref{fig5}c) the centre
region is compressible with $\nu(0)$ slightly larger 
than 2 and with the $n=1$ LL pinned to $\mu^{\star}$, whereas the compressible
regions near the density maxima have $\nu(x)>4$ ($\nu(x_{min})\approx 5$) and
the  $n=2$ LL is  pinned to $\mu^{\star}$. This yields $var\approx \Omega$.
For slightly larger $\Omega$ [(ii), dashed lines in \ref{fig5}b and
\ref{fig5}c], an incompressible strip with filling factor 2 and a local
maximum of $V(x)$ at $x=0$ develops in the centre, while the situation near
the density maxima is unchanged. This leads to  $\Omega <var <2\Omega$.
At still slightly larger $\Omega$ [(iii), dotted lines in \ref{fig5}b and
\ref{fig5}c], the region near the  density maxima is still qualitatively
unchanged, whereas in the centre a compressible region with $\nu(x)<2$
develops, where the  $n=0$ LL  is  pinned to $\mu^{\star}$, leading to 
 $var\approx 2 \Omega$.
As $\Omega$ increases further [(iv) -- (vi)], the situation in the centre
remains qualitatively the same, whereas the filling factor at the density
maxima changes from $\nu(x_{min}) >4$ to $\nu(x_{min}) <4$, so that the
incompressible strips with $\nu(x)=4$ merge [(v), dashed lines in
\ref{fig5}d and \ref{fig5}e] and finally disappear, accompanied by the
disappearance of the potential step across these strips. Thus, we have again 
 $var\approx \Omega$, until for much larger $\Omega$ the maximum filling
 factor becomes smaller than 2 and $var$ goes back to the very small value
 corresponding to linear screening in the lowest Landau level.

In summary, the low-temperature screening properties of a confined
inhomogeneous 2DEG in strong magnetic fields can be understood from a few 
basic facts: Incompressible regions with integer values of the local
filling factor (even integer values for a spin-degenerate 2DEG), and thus
constant electron density, are accompanied by steps of the self-consistent
potential between the adjacent compressible regions, where adjacent Landau
levels are pinned to the constant electrochemical potential, so that  the step
height is $\hbar \omega_c$, the cyclotron energy. The modulation-induced
variation of the self-consistent potential (``variance'') is, therefore,
usually an integer multiple of  $\hbar \omega_c$ and changes by  $\pm \hbar
\omega_c$ if incompressible regions appear or disappear (near local maxima or
minima of the electron density) due to a variation of modulation strength or
magnetic field. Screening of the applied modulation potential breaks down if
the electron density near the density minima becomes so small that no
further redistribution of electrons from the local potential maxima to the
local potential minima is possible.

%%%%%%%%%%%%%%%%%%
\section*{Acknowledgements}
We gratefully acknowledge helpful discussions with E. Ahlswede
and J. Weis, and financial support by the Deutsche Forschungsgemeinschaft, SP
``Quanten-Hall-Systeme'' GE306/4-1.

% This is how we recommend you do the references in LaTeX
\section*{References}
\bibliography{/home/gerha/maloch/zitate}
\bibliographystyle{plain}

%\begin{thebibliography}{9}
%% The next line could be included to match the Word file
%% By default the references are printed smaller in the iopart style
%% \normalsize
%\bibitem{Chklovskii92:4026}
%Chklovskii D B, Shklovskii B I and Glazman L I 1992 
%Phys. Rev. B 46 4026

%\bibitem{Oh97:13519}
%Oh J H and Gerhardts R R 1997
%Phys. Rev. B 56 13519
%\bibitem{Wulf88:4218}
%Wulf U, Gudmundsson V and Gerhardts R R 1988
%Phys. Rev. B 38 4218
%\bibitem{Lier94:7757}
%Lier K and Gerhardts R R 1994
%Phys. Rev. B 50 7757
%\end{thebibliography}

\end{document}